\documentclass{PoS}

\newcommand{\Ord}{\mathcal{O}}
\newcommand{\refc}[1]{(\ref{#1})}

\title{Spectrum of the open QCD flux tube in $d=2+1$ and its effective
string description}

\ShortTitle{Spectrum of the open QCD flux tube in $d=2+1$ and its effective
string description}

\author{\speaker{Bastian B. Brandt} \\
        Institute for Theoretical Physics, University of Regensburg, \\
        93040~Regensburg, Germany \\
        E-mail: \email{bastian.brandt@physik.uni-regensburg.de}}

\abstract{Simulations in lattice gauge theory suggest that the formation
of a flux tube between quark and antiquark leads to quark
confinement. It is conjectured that the infrared behaviour
of the flux tube is governed by an effective string theory
and simulations show good agreement between lattice data and
its predictions. To next-to leading order ($R^{-3}$) in the inverse
$q\bar{q}$ separation $R$ the effective string theory is equivalent
to Nambu-Goto string theory. For the open flux tube in three
dimensions corrections appear at order $R^{-4}$. We compare these predictions
to high-accuracy measurements of the groundstate energy of the flux tube in 3d
$SU(2)$ and $SU(3)$ gauge theory and extract the coefficient of the leading
order boundary term in the effective action.}

\FullConference{The European Physical Society Conference on High Energy Physics
\\
         18-24 July, 2013\\
         Stockholm, Sweden}

\begin{document}

\section{Introduction}

Lattice simulations provide strong evidence that the formation of a flux tube
between a static quark and antiquark is a valid mechanism to describe the
confinement of quarks. In the ultraviolet it is conjectured that the dynamics
of the flux tube is well described by an effective string
theory~\cite{Goddard:1973qh,Nambu:1978bd,Luscher:1980fr,Polyakov:1980ca}. A
simple ansatz for the woldsheet field theory is the action of a free bosonic
string, the Nambu-Goto (NG) action with the spectrum~\cite{Arvis:1983fp}
\begin{equation}
\label{eq:NG-spectrum}
E^{\rm NG}_{n}(R) = \sigma \: R \: \sqrt{ 1 + \frac{2\pi}{\sigma\:R^{2}} \:
\left( n - \frac{1}{24} \: ( d - 2 ) \right) } \;.
\end{equation}
Here $\sigma$ is the string tension and $R$ the $q\bar{q}$ separation. In
$d\neq26$ the quantisation of the NG action suffers from the well known Weyl
anomaly~\cite{Goddard:1973qh}, rendering the interpretation of the NG
action as the low energy effective action for the flux tube in the
ultraviolet region inconsistent.

This consistency problem has triggered the development of alternative frameworks
to write down the effective string action. Historically there are two
main methods, the ansatz by {\it Polchinski} and {\it
Strominger}~\cite{Polchinski:1991ax} and the one by {\it L\"uscher} and {\it
Weisz}~\cite{Luscher:2002qv,Luscher:2004ib} and for a long time it was unknown
how to connect these two. Recently {\it Aharony} and collaborators
performed extensive calculations in both theories, showing that the spectrum is
equivalent in the two theories up to $\Ord(R^{-5})$ and that it includes
corrections to the NG spectrum~\cite{Aharony:2009gg,Aharony:2010db,Aharony:2010cx,
Aharony:2011ga}. Furthermore, they
provided a relation between the two methods based on a generalisation of the
AdS/CFT correspondence for pure gauge theory where the confining string is
a weakly coupled fundamental string moving in a weakly curved
background~\cite{Aharony:2013ipa}. Within this framework the two approaches
correspond to the `conformal' or `static/unitary' gauge, respectively.
Alternative methods to calculate the
spectrum of the confining string have also been proposed
recently~\cite{Dubovsky:2012sh,Caselle:2013dra}, leading to compatible
predictions.

Simulations in pure gauge theories~\footnote{For results see
\cite{Brandt:2009tc,Mykkanen:2012dv} (open spectrum),
\cite{Teper:2009uf,Athenodorou:2010cs,Athenodorou:2011rx} (closed spectrum),
\cite{Pepe:2010na} (width) and the included lists of references.} show striking
agreement with the NG predictions for the energy levels of the open and the
closed flux tube, as well as for the width of the flux tube at its midpoint.
The accordance persists down to surprisingly small values of the $q\bar{q}$
separation where the flux tube cannot be expected to be shaped like a string.
Deviations from NG became visible only recently with increasing precision in the
measurements of the energy levels. For the open string spectrum these deviations
are well described by the predictions from the effective string
theory~\cite{Brandt:2010bw,Billo:2012da}. Similar agreement has also been found
concerning the predictions at finite
temperature~\cite{Bialas:2009pt,Caselle:2011vk} and for other objects in gauge
theories with an effective string description (e.g.~\cite{Caselle:2007yc} and
references therein).

In this proceedings article we present new results concerning the coefficient of
the leading order correction to the NG spectrum of the open string. We
begin by summarising the predictions from the effective string theory before
we discuss briefly the details of the simulations and the extraction of the
flux tube spectrum. We close with the discussion of the results for the
boundary coefficient.

\section{Spectrum of the open confining string}
\label{sec:spect}

In the fundamental effective string theory, discussed in detail
in~\cite{Aharony:2013ipa}, the full NG action is the only `weight
zero'~\footnote{In this framework the `weight' counts the number of derivatives
in the action. The induced metric, with two derivatives, has zero weight per
definition.} term entering the action, so that all additional terms in the
spectrum will appear with respect to eq.~\refc{eq:NG-spectrum}. For the open
string the first correction that appears is a four-derivative boundary
term~\cite{Aharony:2009gg}. The associated energy levels are given
by~\cite{Aharony:2010db}
\begin{equation}
\label{eq:spectrum}
 E^{\rm BC}_n(R) = E^{\rm NG}_n(R) - \bar{b}_2
\;\frac{\pi^3}{\sqrt{\sigma^3}\;R^4}\;\left( 4\;\overline{N}_n +
\frac{d-2}{60} \right) \;,
\end{equation}
where $\overline{N}_n$ is a factor that depends on the eigenstate at energy
level $n$ (for a classification of the states
via charge conjugation C and parity P, $\overline{N}_n=\overline{N}_n^{\rm CP}$
and $\overline{N}_0=0$, $\overline{N}_1=1$, $\overline{N}_2^{++}=2$ and
$\overline{N}_2^{--}=8$; see~\cite{Brandt:2010bw}) and $\bar{b}_2\equiv
\sqrt{\sigma^3}b_2$ is a dimensionless non-universal parameter. The same result
has also been found using an alternative ansatz~\cite{Billo:2012da}.
The next correction is a regular (non-boundary) term appearing at
$\Ord(R^{-5})$ and can be expected to be non-universal as well. However, in
$d=2+1$, the case considered in this study, it vanishes identically.

\section{Simulation details}

\begin{table}
\centering
\small
\vspace*{-3mm}
\begin{tabular}{cc|cccc|ccccc}
$N$ & Lat & $\beta$ & $r_{0}/a$ & $\sqrt{\sigma}\;a$ & $\sqrt{\sigma}r_0$ &
size &
$R/a$ & $t_s$ & $n_t$ & \#meas \\
\hline
\hline
 & & & & & & & & & & \vspace*{-4mm} \\
2 & A & 5.0 & 3.9472(4) & 0.31212(7) & 1.2320(3) & $48^3$ & 2-12 & 2 & 30000 &
1600 \\
 & B & 7.5 & 6.2863(7) & 0.19630(4) & 1.2340(3) & $64^3$ & 2-26 & 4 & 30000 &
1400 \\
 & C & 10.0 & 8.5992(6) & 0.14354(2) & 1.2343(2) & $96^3$ & 2-30 & 6 & 30000 &
2200 \\
\hline
 & & & & & & & & & & \vspace*{-4mm} \\
3 & F & 14.0 & 4.4433(3) & 0.27682(4) & 1.2300(2) & $48^3$ & 2-13 & 2 & 20000 &
1900 \\
 & G & 20.0 & 6.7073(4) & 0.18367(2) & 1.2319(2) & $60^3$ & 3-27 & 4 & 20000 &
2300 \\
\hline
\hline
\end{tabular}
\caption{List of simulation parameters. $R$ denotes the $q\bar{q}$ separation,
$t_s$ is the temporal extent of the L\"uscher-Weisz sublattices and
$n_t$ denotes the number of sublattice updates.}
\label{tab1}
\vspace*{-3mm}
\end{table}

We perform simulations of pure $SU(2)$ and $SU(3)$ gauge theory in 2+1
dimensions using the standard combination of heatbath and overrelaxation
updates. The simulation parameters and lattice sizes are listed in
table~\ref{tab1}.

The focus is on the groundstate potential that can be extracted with high
precision using Polyakov loop correlation functions. For those correlators
excited states are suppressed exponentially with $T$ so that their contribution
to the groundstate signal can usually be neglected. In the following the
energies are extracted via
\begin{equation}
 \label{eq:energies}
 V(R) \equiv E_0(R) = -\frac{1}{T} \ln\left[\left< P^{\ast}(R)\:P(0)
\right> \right] \;,
\end{equation}
where $T$ is the temporal extent of the lattice. The scale is set via the string
tension as defined by the potential from eq.~\refc{eq:spectrum}. For
completeness we also list the result for the Sommer scale and $\sqrt{\sigma}r_0$
in table~\ref{tab2}. The Polyakov loop correlation functions have been measured
using the L\"uscher-Weisz multilevel algorithm~\cite{Luscher:2001up} with the
parameters given in table~\ref{tab1}. We have checked that finite size effects
as discussed in~\cite{Brandt:2010bw} are small for all spatial extents $R$. The
error analysis is done using the jackknife method with 100 bins for all
lattices. We have also checked that varying the number of bins did not change
the error estimates.

\section{Extraction of the boundary coefficient}

\begin{table}
\centering
\small
\vspace*{-3mm}
\begin{tabular}{c|cccc}
 Lat & $R_{\rm min}/a$ & $\bar{b}_2$ & $a V_0$ & $\chi^2/$dof \\
\hline
\hline
 & & & & \vspace*{-4mm} \\
A & 4 & -0.0197 (11)(52) & 0.2148 (2)(3) & 0.1 \\
B & 5 & -0.0210 ( 4)(16) & 0.1740 (1)(2) & 0.1 \\
C & 8 & -0.0231 ( 1)(22) & 0.14500 (4)(11) & 0.2 \\
\hline
 & & & & \vspace*{-4mm} \\
F & 4 & -0.0137 ( 4)(10) & 0.2239 (1)(2) & 0.02 \\
G & 6 & -0.0170 ( 3)( 1) & 0.18166 (5)(1) & 0.02 \\
\hline
\hline
\end{tabular}
\caption{Results for the fitparameters as explained in the text.}
\label{tab2}
\vspace*{-3mm}
\end{table}

\begin{figure}
\centering
\begin{minipage}[c]{0.49\textwidth}
\includegraphics[width=\textwidth]{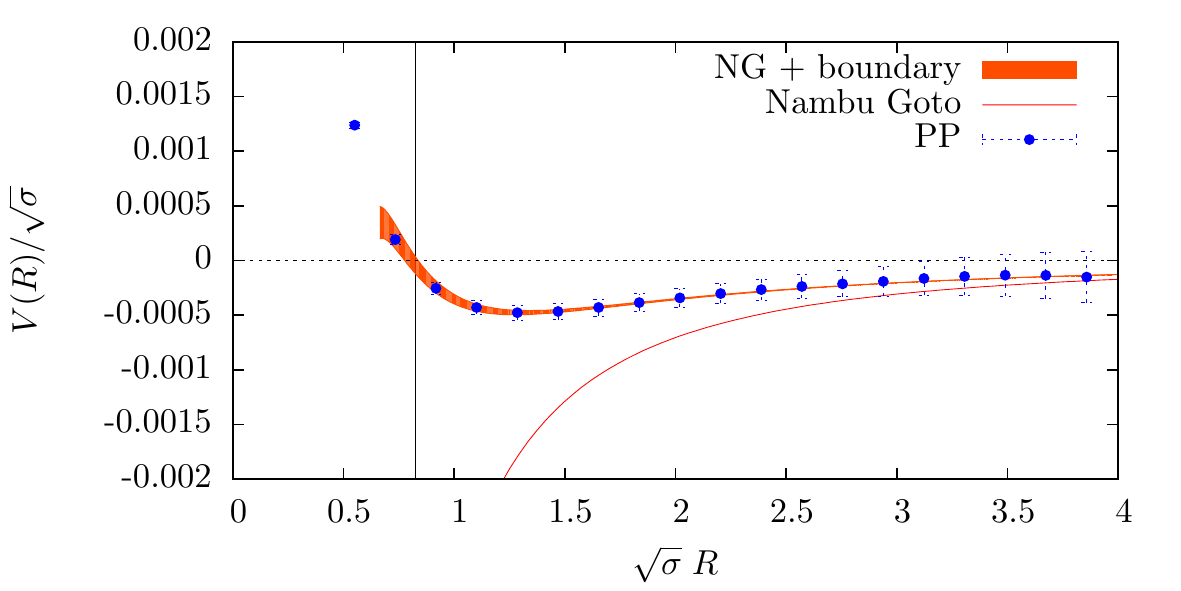}
\end{minipage}
\begin{minipage}[c]{0.49\textwidth}
\includegraphics[width=\textwidth]{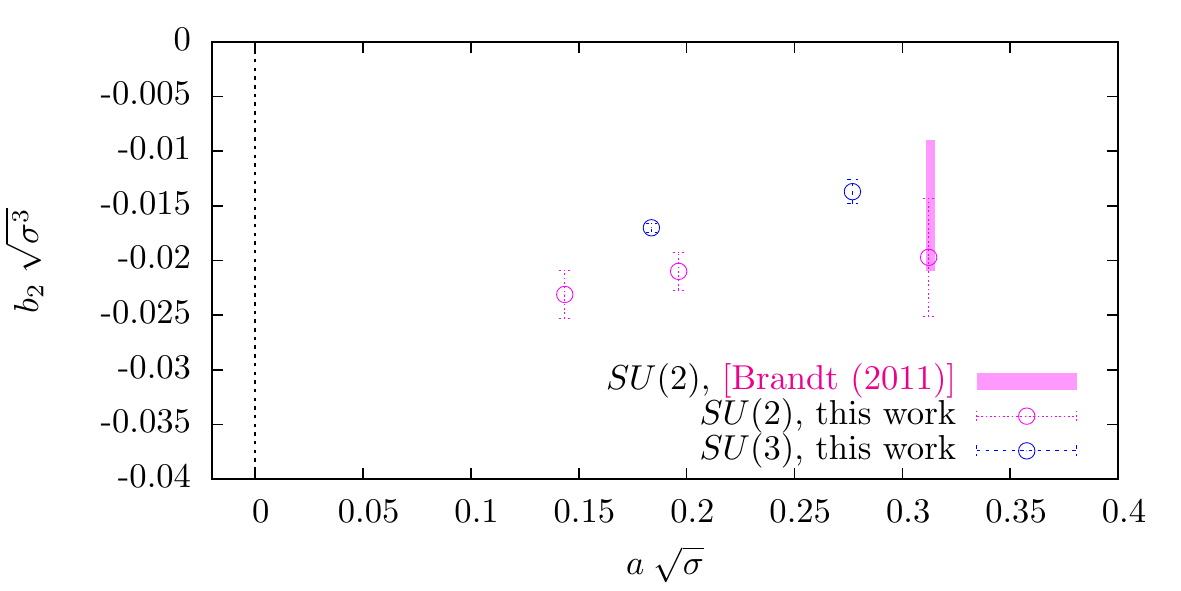}
\end{minipage}
\caption{{\it Left:} Results for the groundstate $q\bar{q}$ potential on
lattice G, normalised as in~\cite{Brandt:2010bw}. The orange band is the
fit result and the vertical line indicates the fitrange. \hspace*{2mm}{\it
Right:} Results for $\bar{b}_2$ versus the lattice spacing in units of the
string tension in comparison to the result from~\cite{Brandt:2010bw}.}
\label{fig1}
\vspace*{-3mm}
\end{figure}

The boundary coefficient has been extracted from the groundstate energies by a
fit to the form given in eq.~\refc{eq:spectrum} with $\sigma$, $\bar{b}_2$ and
an additive normalisation constant $V_0$ as free parameters. The lower cut
$R_{\rm min}$ has been chosen to be the second smallest $q\bar{q}$ separation
for which the fit still provided $\chi^2/$dof values smaller than one and is
listed in table~\ref{tab2} together with the results for $\bar{b}_2$ and $V_0$.
$\sigma$ is given in table~\ref{tab1}. The systematic error associated with this
particular choice (the second error for the quantities in table~\ref{tab2}) is
estimated from the spread of the results obtained by fits with a minimal
$q\bar{q}$ separation $R_{\rm min}\pm1$. Figure~\ref{fig1}
(left) shows the result for the groundstate potential of lattice G in
comparison to the fit and the NG prediction. The plot displays the excellent
agreement between the predictions and the data.

In the case of $SU(2)$ the error of $\bar{b}_2$ is dominated by the
systematic uncertainty. This is different for $SU(3)$, where in
general much less dependence on $R_{\rm min}$ is seen. Furthermore,
$\chi^2/$dof differs by an order of magnitude between the two cases. 
A possible explanation is provided by the fact that the picture of a single
non-interacting flux tube is consistent only in the limit $N\to\infty$. This
suggests that we observe a remnant of the interaction of the flux tube with
the lightest glueballs which perturbs the agreement with the effective string
predictions (see also the discussions
in~\cite{Athenodorou:2010cs,Dalley:2005fw,Lucini:2012gg}).

In figure~\ref{fig1} (right) the results for $\bar{b}_2$ are plotted versus the
lattice spacing in units of the string tension. The dependence on the lattice
spacing is relatively mild in both cases and the results are close together,
indicating only a small $N$-dependence of $\bar{b}_2$. Whether this is also
true in the continuum remains to be seen. A detailed continuum extrapolation
together with comparisons to data for the excited states will be presented in a
future publication.

It would be interesting to compare to the results from $Z(2)$ gauge
theory~\cite{Billo:2012da}. This is difficult, since in~\cite{Billo:2012da} the
coefficient $\bar{b}_2$ has been extracted using the expansion of the NG
energies to $\Ord(R^{-3})$ instead of the full resumed energies. Note, that
extracting $\bar{b}_2$ from an expansion of eq.~\refc{eq:NG-spectrum} to
different orders in $1/R$ even leads to different signs for $\bar{b}_2$, since
the signs of the expansion terms alternate. Following the discussion in
section~\ref{sec:spect} and in~\cite{Aharony:2010db,Athenodorou:2010cs}, starting
from eq.~\refc{eq:spectrum} thus appears to be a proper and unambiguous choice
for the extraction of $\bar{b}_2$.

{\bf Acknowledgments:} I would like to thank M. Panero for the invitation to
give this talk and M. Caselle for discussions. The simulations were done in
parts on the LC2 cluster at the university of Mainz and on the computing
resources of the university of Regensburg.

\end{document}